# A PROPOSED METHOD USING GPU BASED SDO TO OPTIMIZE RETAIL WAREHOUSES

*Associate professor Magnus Bengtsson, PhD, University of Borås, Sweden*
*Associate professor Jonas Waidringer, PhD, University of Borås*

**University of Borås**
*Allégatan 1*
*503 32 Borås*
*+46(0)334354122, +46(0)334354617*
*magnus.bengtsson@hb.se – jonas.waidringer@hb.se*

*Abstract:*
*Research in warehouse optimization has gotten increased attention in the last few years due to e-commerce. The warehouse contains a waste range of different products. Due to the nature of the individual order, it is challenging to plan the picking list to optimize the material flow in the process. There are also challenges in minimizing costs and increasing production capacity, and this complexity can be defined as a multidisciplinary optimization problem with an IDF nature. In recent years the use of parallel computing using GPGPUs has become increasingly popular due to the introduction of CUDA C and accompanying applications in, e.g., Python.*
*In the case study at the company in the field of retail, a case study including a system design optimization (SDO) resulted in an increase in throughput with well over 20% just by clustering different categories and suggesting in which sequence the orders should be picked during a given time frame.*
*The options provided by implementing a distributed high-performance computing network based on GPUs for subsystem optimization have shown to be fruitful in developing a functioning SDO for warehouse optimization. The toolchain can be used for designing new warehouses or evaluating and tuning existing ones.*

1. **Introduction**

The E-commerce market is one of the fastest-growing industries globally, and with the Covid-19 pandemic, the effect is even more distinctive. The southwest parts of Sweden, where Borås and Gothenburg are, hold one of Northern Europe's most significant clusters of e-commerce companies. With e-commerce both the need for warehousing and manpower is large and with that comes some specific difficulties. Today, an e-commerce company can offer a customer anything between 100 to 200,000 products, where orders placed before midnight are processed, packed and sent by 4 pm the same day. It is not unusual to pick around 10,000 items in one day, spread over 4,000 orders and over 200,000 locations in the warehouse.(Kordos, Boryczko et al. 2020) write that a large share of operating costs related to product storage is connected to order picking. Studies have established that about 60% of warehouse operation costs are linked to charges of picking up goods when completing orders. The time required for this operation is a decisive factor in the level of customer satisfaction. The lead time from when an order is placed to the delivery decreases, and short lead times are something that the customer expects.

Order picking is the process that determines how quickly a product can be shipped out to the customer. An inefficient picking process can generate customer dissatisfaction which leads to a loss for the company. (de Koster, Le-Duc et al. 2007) Order picking is the process in warehouse management that requires the most resources. By strategically placing products, you can facilitate order picking and reduce handling time, but another critical element is finding the picker's optimal route.(Silva, Coelho et al. 2020) With the wide variety and quantity of articles, the requirement increases that the company must have efficient processes for stock management so that the goods quickly and easily can be picked without unnecessary lead times.

Due to the high demand for fast deliveries and many different articles, it is hard to optimize current warehouses; therefore, artificial intelligence might be an option. AI is a machine learning algorithm that can help with the recognition of patterns. With the help of AI, you can structure and analyze data in events that today are too complex.(O'Leary 2013) The need to study this problem is also found in society and its welfare and the individual consumer. Logistics is one of three pillars for sound economic development. It has even been seen that logistics play a crucial role in a city's economic growth. (Song, Yeung et al. 2020)

Research in warehouse optimization has gotten increased attention in the last few years due to e-commerce. The warehouse contains a waste range of different products. Due to the nature of the individual order, it is challenging to plan the picking list to optimize the material flow in the process. There are also challenges in minimizing costs and increasing production capacity, and this complexity can be defined as a multidisciplinary optimization problem with an IDF nature (Panos Y. Papalambros 2017). In recent years the use of parallel computing using GPGPUs has become increasingly popular due to the introduction of CUDA C and accompanying applications in, e.g. Python.

Decision-making as a hybrid system is schematically described in Figure 1 below. Scenario simulations will form part of the input to the training of machine learning algorithms. Machine learning algorithms classify information from previous decision bases and are trained to make decisions according to different scenarios—the decision-maker for a tailor-made decision basis that leads to a more optimal conclusion. The various topologies are parallelized and simulated according to the global topology as shown.

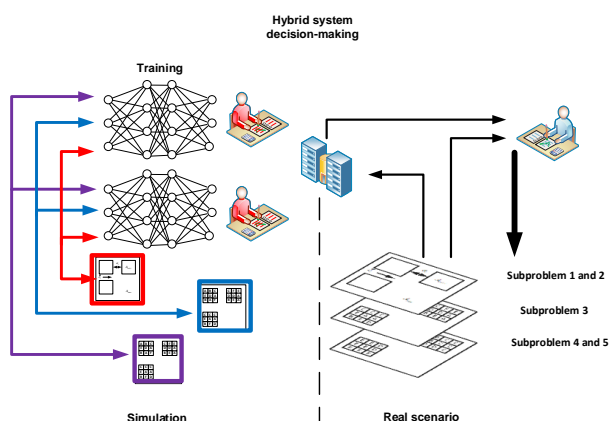

FIGURE 1 HYBRID DECISION MAKING

2. **Frame of reference**

The field of AI, machine learning and dynamic systems optimization can be a subpart within the area of Complex Adaptive Systems (CAS). Complex adaptive systems and AI share a common denominator in Cybernetics. (Howard J. 2019) Both are conceptual at their core and has thus evolved and been more specific as in machine learning algorithms and different approaches such as SDO (Systems Design Optimisation), which this article is focused on.

The combination of new technologies gives us new opportunities to manage any industry. The retail sector is exciting, with large amounts of articles and stakeholders in a dynamic adaptive environment. This is closely related to the core properties of a socioeconomic or socio-technical system such as the retail industry. This can conceptually be described as in the example below in Figure 2 from the transportation and logistics industry, which retail belongs to.

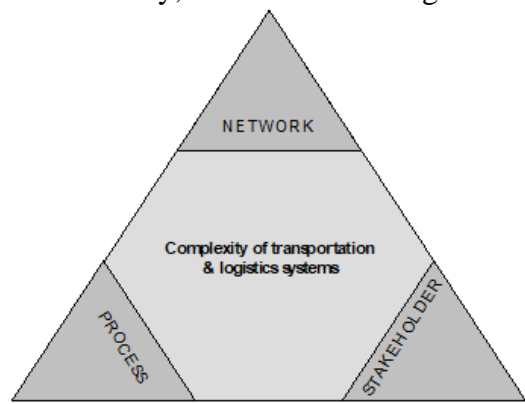

FIGURE 2 CORE PROPERTIES OF TRANSPORTATION AND LOGISTICS SYSTEMS' COMPLEXITY (WAIDRINGER 2001)

Overall the network property is primarily concerned with the design and redesign of any system. The network complexity property is dependent on which type of network/system it is, its overall structure and connectedness in terms of nodes and links. The complexity derived from the network regards the accessibility and availability of the network at hand and the right or" optimal" utilization.

The process perspective is primarily concerned with the operations of an actual setup of a system. The process complexity property depends on which type of process or flow it is, on the activities that build up the process and the function it is performing and the interfaces between these flows. The complexity derived from this property regards the right or "optimal" combination of these activities. The activities can be coupled, parallel or decoupled, i.e. through a buffer in terms of stock or information in the case of the retail industry.

The stakeholder perspective is primarily concerned with management and control of the process and network. The stakeholder complexity property is dependent on which hierarchical level is considered, the decisions made of the different stakeholders and what impact these decisions have. The complexity derived from this property regards the right or "optimal" organizations of these stakeholders to make the decisions as efficiently as possible.

These three properties and interdependencies make up inherently complex systems, where AI and machine learning approaches are deemed appropriate.

The perception of supply networks and logistics systems, such as large warehouses with a heterogenous mix of products as being complex is a well established concept and emphasized by several authors (Bowersox 2002), (Christopher 2016) (Cox 1999) (Lumsden Kenth. R.; Hulthén Lars. A. R.; Waidringer 1998, Tan 2001) (Waidringer 2001) Lendel, Pancikova et al. (2016)). However, the perceived complexity is often derived from an interpretation of logistics systems as being difficult to understand since these systems consist of many parts, relationships, and flows, i.e. they should be heavily reduced and simplified to be dealt with.

It is also necessary to understand the subtle but essential difference between logistics and supply chain management at a conceptual level. Retail and warehouse management is an integral part of both these systems. All flow-related activities can be viewed in an abstract and a physical, or rather from a business and an engineering, perspective. The concept of supply chain management has its origin in the business perspective where transactions are focused and where it is essential to manage and control information about the goods or services to be delivered (Cooper et al, 1997). The business perspective also contains the abstract strategic dimensions of the supply chain (Bechtel and Jayaram, 1997).

The concept of logistics has its origin in the military perspective regarding the physical, operational treatment, and transfer of goods, which focuses on the need for the engineer to find workable solutions to the displacement, handling and storage problems. However, the word logistics in the business community is a much more recent post World War II tradition, which is closely coupled to the growth of the information and communication industry (Sjöstedt, 1994). Thus, the business system represents abstract thinking regarding possible future solutions building on knowledge and extrapolation of historical data, i.e., discounting future developments. In contrast, the engineering system means the actual course of events where everything in principle is operationalized. Altogether, this results in a conceptual figure of the core function of any logistical system here referred to as management of flows, regardless of which flows, see Figure 3.

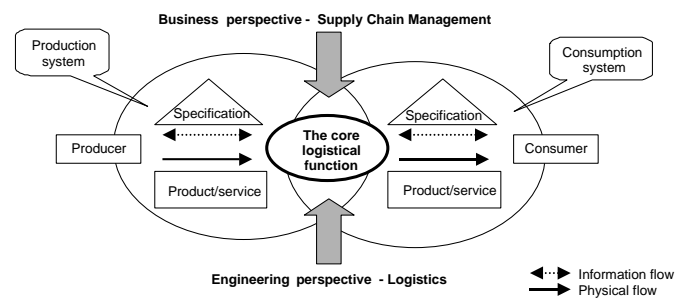

FIGURE 3 THE ROLE OF THE CORE LOGISTICAL FUNCTION IN A VALUE-ADDED INDUSTRIAL SYSTEM

The figure describes how the consumption functions, as part of marketing or other business activities, generate a specification transferred through the logistical process to the producer. First, through the production function, the producer materializes the specification into a tangible product or service, which is brought forward to the consumer through the logistical process. Next, the producer has to judge how much resources have to be used to meet the specification satisfactorily, which basically is a judgment of the market opportunity. In their turn, the consumer judges how well the product or service correlates with their expectations, which is a utility evaluation. (Sjöstedt et al, 2001).

Through SCM and similar concepts, the business system manages and controls the resources in the engineering system where the actual handling of goods and services is done. Thus, the warehouse is connecting the two different systems. Therefore, it can be the essential core logistical function in the supply chain, allowing customers' demands to be met.

### 3. System design optimization

The MDF approach works well if the subsystems are weakly coupled and the analysis is not too computationally expensive. If the system analysis does not converge, there will not be a beneficial outcome from the MDF optimization. The individual disciplinary feasible strategy will have local optimization in each subsystem, and the result cannot guarantee that all subsystems will have possible solutions. IDF has a decision variable that includes both design and coupling variables, which is not the case for MDF, which only has design variables.

IDF problem dimensionality is increased since coupling variables are made design variables (Panos Y. Papalambros 2017). This increase can reduce numerical solution accuracy for large problem sizes (Hulme and Bloebaum 2000). Furthermore, MDF is preferable when the dimension of y is much larger than the dimension of x (Haftka, Sobieszczanski-Sobieski et al. 1992, Alexandrov 2000).

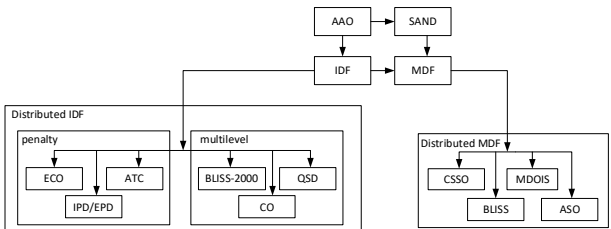
**FIGURE 4 SDO MAP -IDF AND MDF CATEGORIES**

### 4. GPU usage

The main goal with the Mainframe design is that it can be used for solving both MDF and IDF problems using a GPU enabled environment. The means by which multiple GPUs can be organized may vary from application to application. In this paper, we have visualized the communication between the computer nodes with the implementation of TCP servers placed locally at each node. In the case of IDF, we want to control the distribution of calculations via the NHATC algorithm and its data structure for handling design and coupling variables. The approach is chosen to evaluate possible bottlenecks in a specific subprocess.

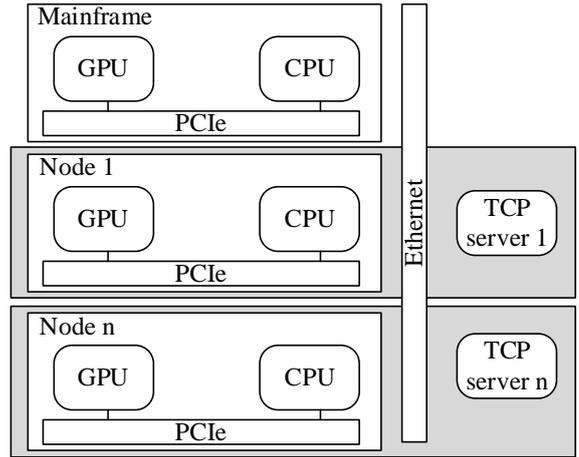
**FIGURE 5 NETWORK REPRESENTATION CPU-GPU-TCP SERVER**

### 5. Case study: Warehouse optimization

Two categories of optimization problems have been analyzed concerning SDO – The first is when the design problem where all sets of subproblems are included in the IDF optimization, see Figure 6 General IDF perspective. The system is complex, and some of the subsystems may have stronger coupling and the degree of freedom is significant. Therefore, this system may need a well-defined set of constraints to find a feasible solution.

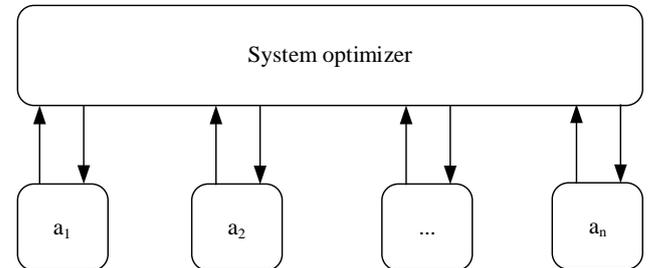
**FIGURE 6 GENERAL IDF PERSPECTIVE**

The reduced process optimization problem includes a subset of subsystems, and some of the excluded subsets act as constraints for the active subsystems, as shown in Figure 7. In this category, the degrees of freedom are lower; hence, finding feasible solutions in a shorter period is more reasonable.

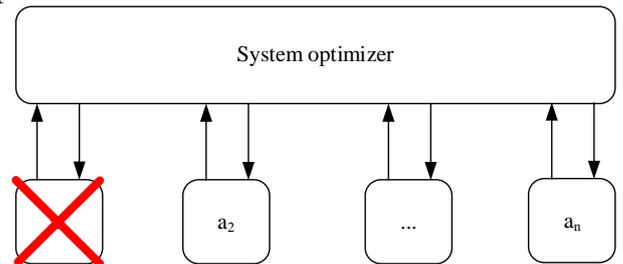
**FIGURE 7 LIMITED SET OF SUBSYSTEMS THAT ARE ACTIVE**

According to the nature of the optimization problem, it can be categorized as a multidisciplinary optimization problem with IDF structure. This paper will apply NHATC to solve the global optimization problem because the local optimization criteria may not be met fully. The local optimized optimization problem is a bounded continuous optimization problem that is optimized using an SQP optimizer. The local optimization problem is assumed to be locally convex.

The placement of the rack will impact the picking time and utilization of the warehouse space. The goal is to achieve as high throughput as possible (picking as many orders per day as possible) and maximizing warehouse space utilization. If a warehouse has many racks, picking orders can increase since the warehouse has a larger buffer for handling larger order volumes. There is a trade-off between packing as many shelves as close to each other and picking time for an order. If the storage consists of long corridors of racks, the picking time will increase; hence the throughput will decrease. The design variables a (distance between shelves) and picking frequency $\delta$ will strongly couple the subsystem 1 and 2.

Suppose the organization of individual products are optimized concerning order history. In that case, subsystems 3 and 5 will be strongly coupled concerning parameter o, the order history and the particular product position x. If the position x is changed, the classification of orders regarding the picklist will change. Furthermore, the picking frequency will affect the degree of parallelism, i.e. how many carriages will be used simultaneously.

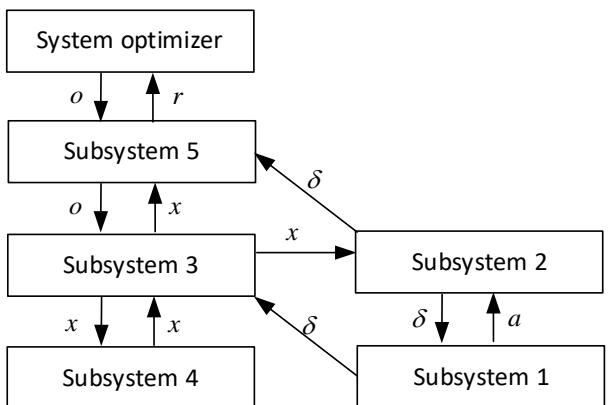

**FIGURE 8 IDF REPRESENTATION OF A RETAIL APPLICATION**

In subsystem 1, the warehouse utilization can be defined as a metric value of the relationship between all rack areas divided by the warehouse floor area as shown in Eq. (1). The objective function has a design variable a, the distance between the racks. In addition, the goal formulation is subject to an inequality constraint consisting of a coupling variable $\delta$ and the design variable $a$. The inequality constraint represents the conflict between large material flow and a small distance between racks, i.e. a small space will limit the number of carriers that can pass a section with width $a$. Finally, the subroutine is optimized using the SQP (Powell 1977) solver coded in CUDA C (Nvidia 2021).

$$\max \frac{A_{ind}}{A_{max}}$$
$$s.t. \quad (1)$$
$$k\overline{\delta}_i - a_i \leq 0$$

Subproblem 2 minimizes the picking route using Dijkstra's algorithm (Dijkstra 1959) to calculate the distance between a and b for a sequence of picking positions. As shown in Eq. (2), the picking route is a variable in the function for calculating picking frequency. Therefore, by evaluating the permutations of different picking sequences, can minimum distance for the picking route be found.

$$\min \sum_{i=1}^{n} d_i$$
$$d_i = subroutine(w_i, a_i, b_i, d_i) \quad (2)$$
$$\delta = f(d)$$

Subproblem 3 optimizes the products placement concerning picking frequency. Due to the nature of the order history, it is beneficial to organize the products' positions concerning how often the products are selected. Furthermore, an unsupervised classification method can determine the optimal placement using, e.g. K-means clustering (Macqueen, 1967). Finally, subroutine 4 will evaluate the possible change in product position regarding, e.g. available empty places, the acceptable cost for reordering the product positions.

In subroutine 5, the individual order represents a specific category of orders. Therefore, using SVM (Cortes and Vapnik 1995) to configure a set of categories that will utilize the output from subroutines 1 to 4 makes it possible to define an optimal picking list. But since the composition of the orders may look different over time, the SVM will need a continuous update making the set of categories in the SVM dynamic. The dynamic update of the SVM is done using a collection of historical data that can be considered to represent the most current order categories. Thus, with time newer data will replace the previous dataset to be used for updating the SVM. The coupling variable $\delta$ will be used to set the constraint for the degree of parallelism, i.e. how many picking cartridges can be used simultaneously.

The individual layers can be analyzed and optimized separately, as visually shown in Figure 9, where, e.g. the rack positions are optimized given a predefined set of picking frequencies (left figure). It can also be route optimization (middle figure) where the rack and product positions are fixed. The figure to the right represents unsupervised clustering.

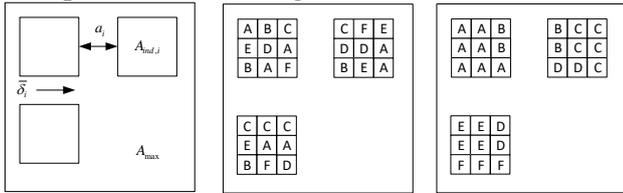

**FIGURE 9 INDIVIDUAL REPRESENTATION OF SUBSYSTEMS LOCAL OPTIMIZATION GOAL.**

This paper argues that using a system design optimization (SDO) view is more beneficial when selecting suitable combinations of subsystems included in the SDO. In Figure 10, individual subsystems are looked upon as parallel systems with a set of mutual variables. The subsystem model may be complex and time-consuming to optimize in a CPU. Furthermore, the joint variables communicated with neighbour systems may require low bandwidth, i.e. there is no bottleneck in information distribution between subsystems. Therefore, we argue that distributing the subsystem optimization to a parallel computing node is beneficial, as shown in Figure 5. Furthermore, the organization of the top-level optimization may be located in a CPU since it has an organizational function to evaluate all information collected from the subsystems.

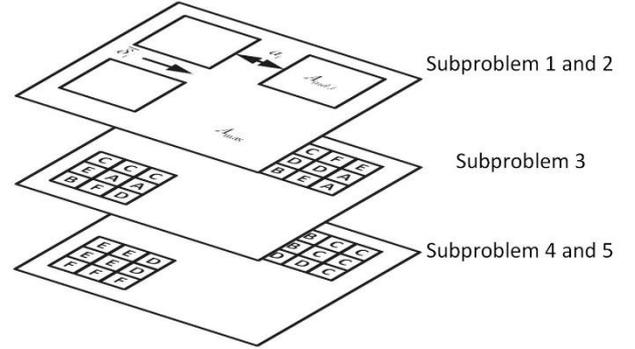

**FIGURE 10 AN MSO PERSPECTIVE OF THE INTERACTION BETWEEN SUBSYSTEMS**

The ATC algorithm (Kim, Rideout et al. 2003) allows communication between parent and child only. Still, since the analyzed case with the warehouse has shown relationships between neighbours as well, it is argued that the system design optimization should be expanded to using the Nonhierarchical ATC (NHATC) (Tosserams, Kokkolaras et al. 2010). In the NHATC, the subproblems have neighbours where targets and responses are communicated (Panos Y. Papalambros 2017), i.e. as the name implies, the order of how information is shared between subsystems has a much larger degree of freedom from a topological perspective.

The general subproblem $P_{ij}$ is formulated as shown in Eq. (3). The target $t$ and response $r$ are embedded in vector $c_{ij}$ (Panos Y. Papalambros 2017). The vector $c_{ij}$ describes the inconsistency between target and response, i.e. the more significant the difference, the more penalizing performance in the objective function since the value of the penalty function $\phi$ increases.

$$\min f_{ij}\left(\bar{x}_{ij}\right)+\phi\left(c_{ij}\right)$$
$$x = \left[ x_{lij}^T, t_{(i+1)k_1}^T, ..., t_{(i+1)k_{nc_{ij}}}^T \right]^T \quad (3)$$
$$s.t. \quad g_{ij}\left(\bar{x}_{ij}\right) \leq 0, \ h_{ij}\left(\bar{x}_{ij}\right) = 0$$

The information flow for the subsystem $P_{ij}$ is shown in Figure 11.

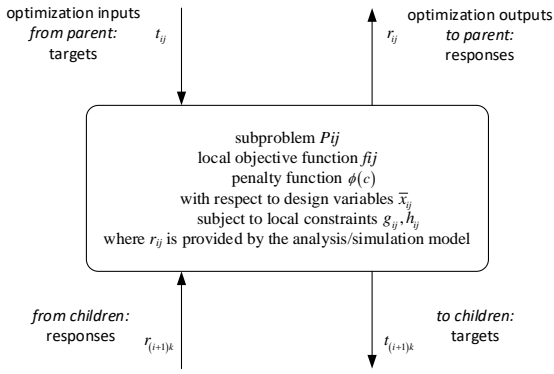

**FIGURE 11 THE DATA FLOW FOR A SUBSYSTEM**

### 6. Results

In the case study at company A, there was no option to reorganize the rack positions initially. Therefore, the initial analysis focused on applying unsupervised classification and route optimization in combination with supervised classification. As a result, the SDO was divided into two sections, where the first section included the unsupervised classification and the route optimization. The second section consisted of supervised classification only. The combined classification and route optimization result showed that the clustering in combination with route optimization could increase the throughput by well over 20% by clustering different categories and suggesting in which sequence the orders should be picked during a given time frame. Unfortunately, in the case of company A, there existed no optimization strategies or tools when initiating the research, so a direct evaluation of improvements compared to existing optimization strategies was not possible. However, from company A:s side, any modification is beneficial given how suitable the change in operation is. Therefore, the first step for company A is to implement a route optimization algorithm in the near time.

### 7. Conclusions

The options provided by implementing a distributed high-performance computing network based on GPUs for subsystem optimization have shown to be fruitful in developing a functioning SDO for warehouse optimization. The toolchain can be used for designing new warehouses or evaluating and tuning existing ones. Even though the results are preliminary, they do indicate that this approach would improve performance for many similar applications. The research in the field of SDO has previously shown how to suggest a suitable optimization strategy. However, it is always the case that the general principle for when to choose to implement an algorithm in a CPU or GPU must be looked upon first. And secondly, the degree of dependency where the SDO is categorized as an IDF of MDF problem. Third, the mutual information shared between subsystems maybe when coupling variables are much larger dimensionally, suggesting that MDF would be more beneficial.

### 8. Nomenclature

| | |
|---|---|
| $r_{ji}$ | response from subproblem $P_j$ to subproblem $P_i$ |
| $t_{ij}$ | target from subproblem $P_j$ to subproblem $P_i$ |
| $x_{li}$ | local variable from $i$th analysis function |
| $x_{si}$ | Shared variable from $i$th analysis function |
| $y_{ij}$ | Coupling variable computed by the $j$th analysis function and required as an input to the $i$th analysis function |
| $\delta$ | Picking frequency |
| $a$ | Distance between racks |
| $g$ | Inequality constraint |
| $h$ | Equality constraint |
| $\phi$ | Penalty function |
| $c$ | Inconsistency vector |
| $o$ | Order $i$ |
| $r$ | Response |
| $t$ | Target |

### 9. Acknowledgements

We would like to thank NVIDIA Corporation for the support of the project within the scope of the NVIDIA Applied Research Accelerator Program. Furthermore, we would like to thank VGR (Region Västra Götaland) for funding the research and the companies involved and the students whose theses were an integral part of the research.

### 10. References

Alexandrov, N. M., Lewis, R. M. (2000). Algorithmic Perspectives on Problem Formulations in MDO. 8th AIAA/USAF/NASA/ISSMO Symposium on Multidisciplinary

Analysis & Optimization. 6-8 September 2000 / Long Beach, CA.

Bowersox, D. J., David J. Closs, Bixby M. Cooper (2002). Supply Chain Logistics Management, Institute for Operations Research and the Management Sciences. **33**: 79.

Christopher, M. (2016). Logistics & supply chain management. Harlow, United Kingdom, Pearson Education.

Cortes, C. and V. Vapnik (1995). "Support-vector networks." Machine learning **20**(3): 273-297.

Cox, A. (1999). "A research agenda for supply chain and business management thinking." Supply chain management **4**(4): 209-212.

de Koster, R., T. Le-Duc and K. J. Roodbergen (2007). "Design and control of warehouse order picking: A literature review." European journal of operational research **182**(2): 481-501.

Dijkstra, E. W. (1959). "A note on two problems in connexion with graphs." Numerische Mathematik **1**(1): 269-271.

Haftka, R. T., J. Sobieszczanski-Sobieski and S. L. Padula (1992). "On options for interdisciplinary analysis and design optimization." Structural Optimization **4**(2): 65-74.

Hulme, K. F. and C. L. Bloebaum (2000). "A simulation-based comparison of multidisciplinary design optimization solution strategies using CASCADE." Structural and multidisciplinary optimization **19**(1): 17-35.

Kim, H. M., D. G. Rideout, P. Y. Papalambros and J. L. Stein (2003). "Analytical target cascading in automotive vehicle design." JOURNAL OF MECHANICAL DESIGN **125**(3): 481-489.

Kordos, M., J. Boryczko, M. Blachnik and S. Golak (2020). "Optimization of Warehouse Operations with Genetic Algorithms." Applied sciences **10**(14): 4817.

Lendel, V., L. Pancikova and L. Falat (2016). Advanced Predictive Methods of Artificial Intelligence in Intelligent Transport Systems, Cham, Springer International Publishing.

Lumsden Kenth. R.; Hulthén Lars. A. R.; Waidringer, J. (1998). Outline for a Conceptual Framework on Complexity in Logistic Systems. Opening markets for Logistics, the Annual Conference for Nordic Researchers in Logistics - 10th NOFOMA 98, Helsinki, Finland, Finnish Association of Logistics.

Nvidia. (2021). "Nvidia CUDA Home Page." Retrieved 14 october 2021, from https://developer.nvidia.com/cuda-zone.

O'Leary, D. (2013). "Artificial Intelligence and Big Data." Intelligent Systems, IEEE **28**: 96-99.

Panos Y. Papalambros, D. J. W. (2017). Principles of optimal design, Cambridge.

Powell, M. J. D. (1977). A fast algorithm for nonlinearly constrained optimization calculations. Numerical Analysis, Dundee 1977. New York, 1978, Springer-Verlag.

Silva, A., L. C. Coelho, M. Darvish and J. Renaud (2020). "Integrating storage location and order picking problems in warehouse planning." Transportation research. Part E, Logistics and transportation review **140**: 102003.

Song, Y., G. Yeung, D. Zhu, L. Zhang, Y. Xu and L. Zhang (2020). "Efficiency of logistics land use: The case of Yangtze River Economic Belt in China, 2000–2017." Journal of transport geography **88**: 102851.

Tan, K. C. (2001). "A framework of supply chain management literature." European journal of purchasing & supply management **7**(1): 39-48.

Tosserams, S., M. Kokkolaras, L. F. P. Etman and J. E. Rooda (2010). "A Nonhierarchical Formulation of Analytical Target Cascading." JOURNAL OF MECHANICAL DESIGN **132**(5).

Waidringer, J. (2001). Complexity in transportation and logistics systems: A conceptual approach to modelling and analysis Doctoral, Report 52, Chalmers University of Technology.